\documentclass[a4paper,journal,final]{IEEEtran}

\usepackage[utf8]{inputenc}
\usepackage[T1]{fontenc}
\usepackage[ngerman,english]{babel}
\usepackage{microtype}
\usepackage{blindtext}
\usepackage{csquotes}
\usepackage{booktabs}
\usepackage{tikz}

\newcommand{\mycitestyle}{ieee}
\IfFileExists{ieee-comp.cbx}{
	\renewcommand{\mycitestyle}{ieee-comp}
}{}
\usepackage[style=ieee,
	citestyle=\mycitestyle,
	sorting=none,
	maxbibnames=6,
	minbibnames=1,
	backend=biber,
]{biblatex}
\defbibheading{ieeerefheading}[\refname]{\section*{References}\footnotesize\vskip 0.3\baselineskip plus 0.1\baselineskip minus 0.1\baselineskip }

\AtEveryBibitem{%
	\iffieldequalstr{entrykey}{grafana}
		{\renewbibmacro*{finentry}{\finentry\newpage}}
		{}%
}

\usepackage{graphicx}
\graphicspath{{../../fig/}{../../build/fig}}

\usepackage[acronym]{glossaries}
\newglossaryentry{usb}
{
	name=USB,
	description={USB}
}

\newglossaryentry{1gbe}
{
	name=\mbox{1\,GbE},
	description={Gigabit Ethernet}
}

\newglossaryentry{10gbe}
{
	name=\mbox{10\,GbE},
	description={10 Gigabit Ethernet}
}

\newglossaryentry{40gbe}
{
	name=\mbox{40\,GbE},
	description={40 Gigabit Ethernet}
}

\newglossaryentry{100gbe}
{
	name=\mbox{100\,GbE},
	description={100 Gigabit Ethernet}
}

\newglossaryentry{vlan}
{
	name=VLAN,
	description={Virtual LAN}
}

\newglossaryentry{asic}
{
	name=ASIC,
	description={application specific integrated circuit}
}

\newglossaryentry{bss}
{
	name=Brain\-ScaleS,
	description={BrainScaleS}
}

\newglossaryentry{bss2}
{
	name=Brain\-\mbox{ScaleS-2},
	description={BrainScaleS-2}
}

\newglossaryentry{cmos}
{
	name=CMOS,
	description={complementary mos}
}

\newglossaryentry{fpga}
{
	name=FPGA,
	description={field programmable gate array}
}

\newglossaryentry{ipmi}
{
	name=IPMI,
	description={Intelligent Platform Management Interface}
}

\newglossaryentry{eda}
{
	name=EDA,
	description={electronic design automation}
}

\newglossaryentry{rtl}
{
	name=RTL,
	description={register transfer logic}
}

\newglossaryentry{icmp}
{
	name=ICMP,
	description={internet control message protocol}
}

\newglossaryentry{jtag}
{
	name=JTAG,
	description={Joint Test Action Group}
}

\newglossaryentry{i2c}
{
	name=\mbox{I\textsuperscript{2}C},
	description={inter-integrated circuit}
}

\newglossaryentry{arp}
{
	name=ARP,
	description={address resolution protocol}
}

\newglossaryentry{sram}
{
	name=SRAM,
	description={static random access memory}
}

\newglossaryentry{snn}
{
	name=spiking neural network,
	description={spiking neural network}
}

\newglossaryentry{fte}
{
	name=full-time equivalent,
	description={full-time equivalent}
}

\newacronym{einc}{EINC}{European Institute for Neuromorphic Computing}
\newacronym{cicd}{\mbox{CI/CD}}{continuous integration and delivery}
\newacronym{pcb}{PCB}{printed circuit board}
\newacronym{soc}{SoC}{system-on-chip}
\newacronym{trl}{TRL}{technology readiness level}
 
\usepackage{mathastext}

\usepackage{siunitx}
\DeclareSIUnit\year{yr}

\usepackage{wasysym}
\usepackage{tipa}

\newcommand{\manualfirstgls}[1]{\acrfull{#1}\glsunset{#1}}

\PassOptionsToPackage{hyphens}{url}
\usepackage[
	hidelinks,
	pdftitle={},
	pdfauthor={},
]{hyperref}
\usepackage[capitalise]{cleveref}

\newcounter{lesson}[section]

\crefname{lesson}{Lesson}{Lessons}
\newcommand{\lesson}[1]{\paragraph{#1}\refstepcounter{lesson}}

\title{Sustainable operation of research infrastructure\\ for novel computing
}

\author{Yannik Stradmann\IEEEauthorrefmark{1},
	Joscha Ilmberger\IEEEauthorrefmark{1},
	Eric Müller,
	and
	Johannes Schemmel

	\thanks{\IEEEauthorrefmark{1}\,These authors contributed equally to this work, listed in random order.
	}
	\thanks{Yannik Stradmann, Joscha Ilmberger and Johannes Schemmel are with the Institute of Computer Engineering, Heidelberg University, 69120 Heidelberg, Germany.
	Eric Müller is with the Kirchhoff-Institute for Physics, Heidelberg University, 69120 Heidelberg, Germany.
	}
	\thanks{Correspondence: Y.~Stradmann (yannik.stradmann@ziti.uni-heidelberg.de).
	}
	\thanks{The presented work has received funding from
	the EC Horizon 2020 Framework Programme under grant agreements 720270 (HBP SGA1), 785907 (HBP SGA2), 945539 (HBP SGA3) and Horizon Europe grant agreement 101147319 (EBRAINS 2.0),
	and
	the \foreignlanguage{ngerman}{Deutsche Forschungsgemeinschaft} (DFG, German Research Foundation) under Germany's Excellence Strategy EXC 2181/1-390900948 (Heidelberg STRUCTURES Excellence Cluster).
	}
}

\newcommand\submittedtext{%
	\footnotesize This work has been submitted to the IEEE for possible publication. Copyright may be transferred without notice, after which this version may no longer be accessible.}

\newcommand\submittednotice{%
	\begin{tikzpicture}[remember picture,overlay]
	\node[anchor=south,yshift=30pt] at (current page.south) {\fbox{\parbox{\dimexpr0.6\textwidth-\fboxsep-\fboxrule\relax}{\centering\submittedtext}}};
	\end{tikzpicture}%
}

\begin{document}

\maketitle

\submittednotice

\begin{abstract}
	Novel compute systems are an emerging research topic, aiming towards building next-generation compute platforms.
For these systems to thrive, they need to be provided as research infrastructure to allow acceptance and usage by a large community.
By the example of the neuromorphic Brain\-\mbox{ScaleS-2} system, we showcase the transformation from a laboratory setup to a publicly available research platform.
It is embedded into a purpose-built institute, tightly coupling a conventional cluster with novel compute hardware.
Our key contribution is the \emph{sustainable operation} of such a research system:
automation, integration, and resource separation enable maintainability, stable operation, and continuity in a fluctuating academic environment.
We implement these concepts by adopting established best practices, such as network infrastructure with strong resource separation, 19-inch rack units, the integration of physical and simulated hardware into \mbox{CI/CD} processes, and system monitoring.
We evaluate six years of platform operation comprising \num{3718} production deployments and an average of \num{198} hardware \mbox{CI/CD} runs per day, with the integration of simulated and physical hardware increasing the full-stack software line coverage from 57\% to 79\%.
We conclude with an explicit collection of lessons learned during the transformation from single-user laboratory setups to a publicly accessible research platform that, since 2023, has served more than 520 individual researchers.
 \end{abstract}

\begin{IEEEkeywords}
	research infrastructure,
analog computing,
platform operation,
continuous integration
 \end{IEEEkeywords}

\section{Introduction}\label{sec:introduction}

One prominent example for novel computing approaches is neuromorphic hardware, which aims to mimic data processing strategies found in neurobiology.
Taking inspiration from the sparsity in communication found in such spiking neural systems, many of these devices target ultra-low-power edge applications~\parencite{frenkel2019morphic,moradi2018dynaps,richter2024dynapse2,yao2024spike,pei2019tianjic,davies2018loihi,wan2022compute,neckar2018braindrop}.
As such, they are usually developed and deployed as standalone \glspl{soc}.
In contrast, large-scale digital systems have been put forward to accelerate compute-intensive workloads and facilitate research in cognitive neurosciences~\parencite{davies2018loihi,Khan2008,gonzales2024spinnaker,merolla2014million,benjamin2014neurogrid,park2016hierarchical}.
Combining both approaches, the \gls{bss2} platform embeds an analog neural network core into digital periphery.
It targets biologically inspired multi-timescale online-learning experiments and therefore emulates neuronal dynamics with a speed-up factor of \num{1000} compared to biological real-time~\parencite{pehle2022brainscales2}.

\begin{figure}[t]
	\centerline{\includegraphics[width=0.9\columnwidth]{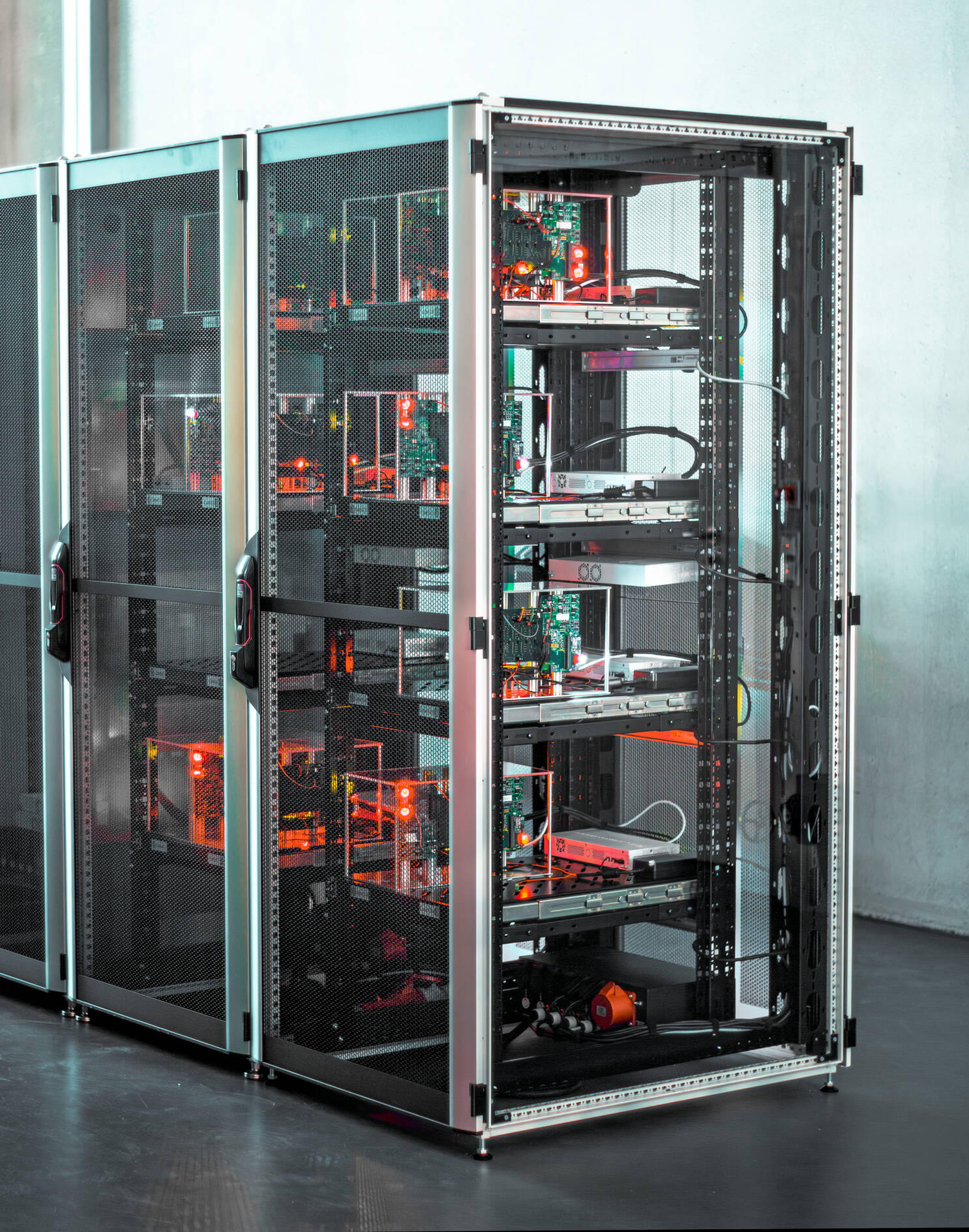}}
	\caption{
		\Gls{bss2} platform in the machine hall of the \acrlong{einc}.
		Two rack cabinets house up to 16 systems, each containing two \glspl{asic}.
		They are placed on retractable drawers to allow maintenance access to individual systems while installed.
		Each of these contains all components for operation, requiring only mains supply and a single Ethernet uplink.
		Both are supplied using local cable passages through the floor to the server room beneath (not visible in the picture).
		\vspace{-3mm}
	}
	\label{fig:bss2_racks_lq}
\end{figure}

Prior to this work, \gls{bss2} was operated as a laboratory system for initial hardware testing and commissioning.
As such, its components were set up as open benchtop devices with immediate access for connecting measurement equipment.
In broader terms, we define the class of \emph{laboratory setups} as those that are described by a selection of the following attributes:
lack of mechanical and electrical integration,
frequent manual interaction---or even required attendance---by skilled users,
inconsistent configuration over time,
no redundancy,
no uptime target,
manual deployment of updates,
and
lack of documentation.

\begin{figure*}[t]
	\centerline{\includegraphics[width=\textwidth]{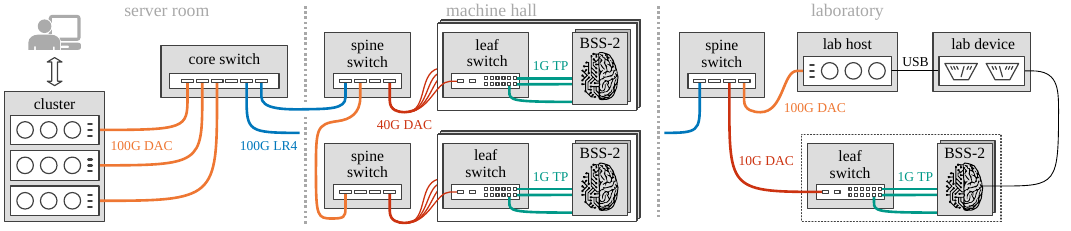}}
	\caption{
		Network infrastructure of the \gls{bss2} (BSS-2) systems, as installed at the \acrfull{einc}.
		The installation is distributed across three floors:
		Users execute experiment code on a compute cluster in the basement, where all nodes are equipped with a dedicated \gls{100gbe} uplink for experiment traffic.
		They connect to a \gls{100gbe} core switch, which in turn provides connectivity to the machine hall (production systems) and laboratories (test systems) via fiber-optic connections.
		In both cases, each rack has a small \gls{100gbe} spine switch, splitting to multiple \gls{10gbe} leaf switches.
		Up until this point, all network traffic consists of multiple \glspl{vlan} dedicated to experiment and management data.
		The leaf switch finally provides untagged \gls{1gbe} connections to the \gls{bss2} setups.
	}
	\label{fig:network}
	\vspace{-3mm}
\end{figure*}
 
This paper presents the transformation of the \gls{bss2} laboratory setup into a mature and versatile research platform for the interdisciplinary fields of cognitive neurosciences and machine learning.
To achieve this goal, we focus on \emph{sustainable operation}: maintaining a usable platform over its lifetime despite changing hardware, software, users, and personnel.

We embed this platform into a purpose-built research building, which couples conventional with novel computing through selected practices of data center design.
However, unlike commercial facilities, the platform operates in an academic setting that permits less stringent availability targets and more relaxed security constraints:
In general, local users are considered non-malicious and have management access to the platform.
Yet they may lack experience or familiarity with established usage conventions and may therefore unintentionally interfere with other users or shared resources.
Consequently, the primary threats considered are accidental misuse, misconfiguration, resource interference, and other unintended actions rather than deliberate attacks.
Throughout this manuscript, we will describe the various mitigation strategies employed to retain stable operation within this environment.
In contrast to local usage, public hardware access relies on a dedicated endpoint that is restricted to experiment data and requires authentication through partner organizations.

Earlier work describes the \gls{bss2} compute system and its user-facing software~\parencite{pehle2022brainscales2,mueller2022scalable}.
In this manuscript, we present the physical infrastructure for operating it as one of the world's largest analog compute systems (\cref{fig:bss2_racks_lq}):
We introduce the Ethernet-based network topology, the composition of the setup, continuous integration workflows, and health monitoring.
Finally, we share lessons learned during the multi-year efforts of operating this infrastructure in an academic setting.

Our key contribution is the adoption of established best practices to enable operational sustainability of an analog, large-scale research system over its full life cycle.
They foster frictionless interplay among a diverse and changing group of platform users and maintainers through high automation, integration, and resource separation.
 \section{Network infrastructure at EINC}\label{sec:network}

The \manualfirstgls{einc} is purpose-built to host research on novel compute systems at large scale.
Across four floors, it houses a state-of-the-art server room, a machine hall, and laboratories:
In its maximum designed configuration, the server room can accommodate \num{63} racks with \SI{1.5}{\mega\watt} \enquote{indirect free cooling} capacity.
The conventional compute cluster in operation consists of \num{26} nodes spread across three racks with a total installed power budget of \SI{30}{\kilo\watt}.
Users execute neuromorphic experiments on the \gls{bss2} platform via this cluster (\cref{fig:network}).
It orchestrates hardware access and transfers data to and from the systems, each utilizing a \gls{1gbe} link.
A core switch (HPE FM3032Q) interconnects all cluster nodes with the neuromorphic hardware in the machine hall and laboratories via \gls{100gbe}.
This dedicated physical network is segmented into multiple IEEE~\mbox{802.1Q} \glspl{vlan} for experiment and different categories of management data:
\begin{enumerate}
	\item Infrastructure management (power distribution, network)\label{vlan:management-infrastructure}
	\item Cluster node management (\gls{ipmi})\label{vlan:management-clusternodes}
	\item \Gls{bss2} system management\label{vlan:management-bss2}
	\item Experiment data to/from \gls{bss2} systems\label{vlan:experiment-bss2}
\end{enumerate}
Together with broadcast rate-limiting (10\%) and weighted queuing disciplines (mapping VLANs to distinct hardware traffic classes with dedicated scheduler (``high'') weights, IEEE~\mbox{802.1Qaz} Enhanced Transmission Selection (ETS) standard), this separation ensures guaranteed proportional bandwidth (management: 4, experiment traffic: 3) for management access in case of unintended behavior of any experiment device, such as denial-of-service.

Network traffic from the server room is distributed to the machine hall (production systems) and laboratories (test systems and maintenance) using fiber-optic connections.
In both cases, it is further distributed through \gls{100gbe} spine switches (MikroTik CRS504-4XQ-IN) to leaf switches (MikroTik CRS326-24G-2S) via \gls{40gbe}-to-4$\times$\gls{10gbe} breakout cables.
Here, the \glspl{vlan} terminate at untagged \gls{1gbe} ports, to which individual \gls{bss2} systems connect.
They comprise one management controller (\gls{vlan}\,\ref{vlan:management-bss2}) and two \glspl{asic}, which communicate experiment data via \glspl{fpga} (\gls{vlan}\,\ref{vlan:experiment-bss2}).

Assuming \gls{1gbe} line-rate loading for each \gls{bss2} setup during an experiment, the sustained maximum data rate per leaf switch (\gls{10gbe} uplink, red) will not exceed \SI{4}{\giga\bit\per\second}.
For the full system (\gls{100gbe} uplink, blue), eight leaf nodes across two racks contribute to a maximum sustained rate of \SI{32}{\giga\bit\per\second}.
Both values are well below the network's capacity, therefore reducing bandwidth-related influence on the latency during the application-dependent tight control- and data loops between accelerators and software.
Thus, the chosen network hierarchy supports concurrent and independent operation of all neuromorphic systems, with headroom for future extensions.
We verify this assumption by independently loading the four \glspl{fpga} per drawer with synthetic data on a custom transport layer.
\Cref{fig:measurements}E,F shows the distribution of the achieved latencies and the empirical cumulative distribution function (eCDF) of the bandwidth measured during \SI{1}{\minute} intervals, for different packet sizes.
Both the latency and bandwidth do not show practically relevant degradation under the additional load on the same leaf switch.
As expected for a network with low utilization, no packet loss was observed throughout more than \num{200000} \Gls{icmp} requests across all load scenarios.
For maximum Ethernet frame size (\SI{1522}{\byte}), the theoretical maximum transport layer payload rate of \SI{949}{\mega\bit\per\second} can be sustained across the \gls{1gbe} link towards the \gls{fpga}.

\begin{figure}[t]
	\centerline{\includegraphics[width=0.9\columnwidth]{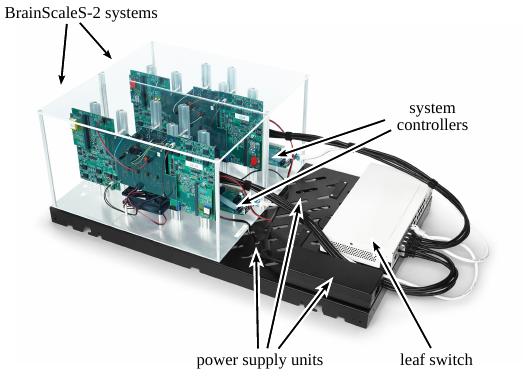}}
	\caption[A single drawer of the \gls{bss2} platform.]{
		A single drawer of the \gls{bss2} platform, containing two systems\footnotemark[1] and all necessary periphery for operation from mains power and a single \gls{10gbe} uplink.
This includes ARM-based system controllers, three power supply units and the leaf switch also shown in \cref{fig:network}.
	}
	\vspace{-2mm}
	\label{fig:bss2_drawer}
\end{figure}

\footnotetext[1]{The shown \gls{bss2} systems have been developed in collaboration with the Chair of Highly-Parallel VLSI Systems and Neuro-Microelectronics at Technische Universität Dresden.
}
\setcounter{footnote}{1}

In addition to this common infrastructure, the laboratories are equipped with remote cluster nodes that accommodate measurement equipment without networking capabilities, such as \gls{usb} devices.
This setup enables a location-agnostic operation of the \gls{bss2} systems in production and during maintenance.
 \section{Systems}\label{sec:systems}

We house all \gls{bss2} systems in conventional 19-inch rack cabinets, which allows easy integration with off-the-shelf components for power distribution and networking.
Since the neuromorphic systems were originally developed for laboratory use, their physical form factor does not comply with this standard.
We therefore mount them on retractable drawers, which provide access to an individual system while the drawer remains installed.
\Cref{fig:bss2_drawer} shows such a unit, where---in addition to two \gls{bss2} systems---all components necessary for operation from mains power and a single network uplink are placed.
Specifically, we provide \SI{12}{\volt} DC power through a single AC adapter per drawer, which feeds all \glspl{fpga}, digital periphery, system controllers, and cooling fans.
The \gls{asic} supplies are derived from a separate \SI{6}{\volt} AC/DC adapter per system to reduce coupling of digital noise into the analog circuits.
We measure less than \SI{120}{\watt} wall-plug power per fully equipped drawer.
As they are spaced in 9\,U intervals, the power density of this setup is very low and no active rack-level cooling is required.
Each analog \gls{asic} contributes less than \SI{1}{\watt}, with the open-frame arrangement leading to neglectable load-dependent thermal crosstalk.

Data connectivity is established through a single \gls{10gbe} uplink, which provides multiple separate \glspl{vlan}.
The managed Ethernet leaf switch exposes these virtual networks as untagged ports for the experiment systems and their ARM-based system controllers.

\label{feat:raspi}
This controller has been inherited from the previous generation of \gls{bss}~\parencite{schmidt2023commissioning}.
It provides \gls{i2c} and \gls{jtag} connections to the associated neuromorphic system for power management and configuration.
It contributes to the collection of \num{29} health metrics, covering supply voltages and currents, temperatures, network reachability and response time, controller status, and the configured \gls{fpga} revision.
This data is streamed to a database over a dedicated Ethernet connection (see \cref{sec:monitoring}).
Over the same interface, users can remotely access system management functionality and \gls{jtag} connections to all \glspl{fpga}.
 \begin{figure*}[t]
	\centerline{\includegraphics[width=0.9\textwidth]{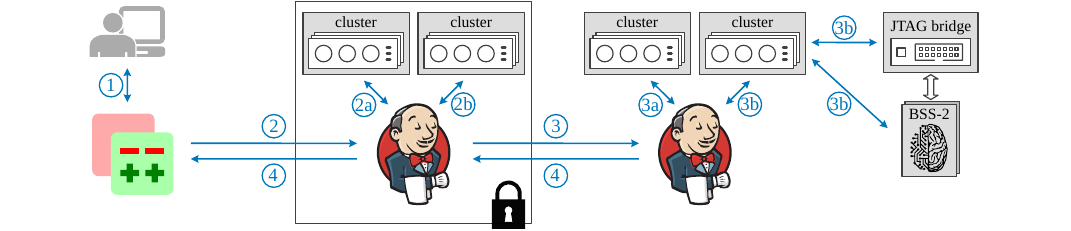}}
	\caption{
		CI/CD workflow for proposed changes to the \gls{bss2} software and hardware configuration.
		Developers upload their patches to a Gerrit instance for code review~\parencite{gerrit}.
		Each upload triggers a build on a security-hardened Jenkins~\parencite{jenkins} instance with access to servers containing commercial \gls{eda} tools.
		These run \gls{rtl} simulations of the combined system and build a bitfile.
		Once the latter is available in binary form, a second Jenkins instance executes downstream tests using this staging bitfile.
		To do so, it builds the \gls{bss2} software stack---including changes the original patch might require---, configures the staging bitfile on a test setup and executes the tests.
		The results will be propagated back to the original patch submitter and be available next to human code-review comments.
	}
	\label{fig:fpga_ci_flow}
\end{figure*}

\section{Platform Integration into CI/CD}\label{sec:continuous-integration}

To support stable operation of the \gls{bss2} platform, we employ a workflow based on \manualfirstgls{cicd}.
In addition to human review, any change to software or hardware components proposed by a developer is automatically verified in simulation and on the hardware systems.
As such, the \gls{bss2} platform itself is tightly integrated into the \gls{cicd} workflow.\footnote{
	Source code available: \url{https://github.com/electronicvisions/jenlib}
}
We demonstrate the steps involved by the example of an in-review software modification that requires an update of the system's \gls{fpga} design (\cref{fig:fpga_ci_flow})
\begin{enumerate}
	\item
	The developer modifies the \gls{rtl} code of the \gls{fpga} locally.
	They indicate the co-dependent software change in the commit message and push the patch to the Gerrit code review system.
\item
	Gerrit notifies a security-hardened Jenkins instance with access to commercial \gls{eda} tools about this proposed changeset.
	This triggers multiple jobs in parallel:
	\begin{enumerate}
		\item \label{cistep:simtests}
		\Gls{rtl}-based simulations of the \gls{bss2} system (\gls{asic} and modified \gls{fpga}).
		The executed test suite includes all referenced software modifications.
		\item \label{cistep:fpgabuild}
		Build of the \gls{fpga} design, resulting in a binary hardware configuration.
		This file, together with metadata, is distributed to a cluster-wide staging area.
	\end{enumerate}
\item
	Once step~\ref{cistep:fpgabuild} has completed, a second---less restricted---Jenkins instance is triggered to perform downstream verification on the physical hardware systems:
	\begin{enumerate}
		\item
		It builds the \gls{bss2} software stack~\parencite{mueller2022scalable} together with test suites ranging from transport layer validation to full-stack experiments---again, including the referenced software modifications.
		\item \label{cistep:hwtests}
		One out of four reserved \gls{bss2} hardware setups is allocated at random, and the staged configuration from step~\ref{cistep:fpgabuild} is loaded into the \gls{fpga} as volatile configuration.
		Subsequently, the prepared test suites are executed on the hardware system.
	\end{enumerate}
\item
	All test results are aggregated and reported to the Gerrit instance in the form of a mandatory vote.
	\item
	After the changeset has additionally been approved by a human reviewer and submitted by the author, a release build is run and delivered as the new stable version.
\end{enumerate}

The strong isolation of individual system components described in \cref{sec:network} as well as the runtime reconfiguration of the \gls{fpga} allows for verification of work-in-progress changesets that may introduce unintended behavior.

\begin{figure*}[t]
	\centerline{\includegraphics[width=\textwidth]{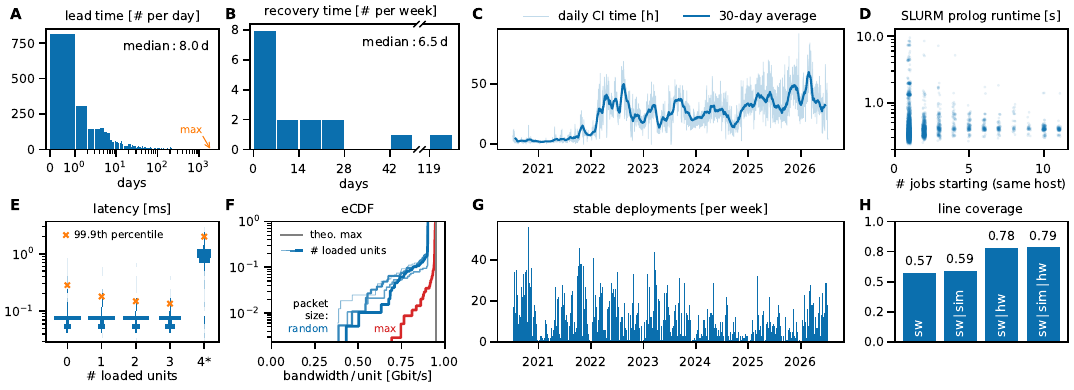}}
	\caption{(A) Distribution of lead times for deployed changes.
		Each change is measured from uploading the initial commit to its stable release.
		(B) Distribution of recovery times for reverted deployments.
		Data includes non-critical rollbacks.
		(C) Time course of compute time spent on neuromorphic hardware for CI (sum over all systems).
		(D) Upper bound for the startup overhead of the job-specific firewall rules described in \cref{feat:firewall}.
		We correlate the full SLURM prolog runtime with the number of concurrently starting jobs on the same host, a proxy for prolog-induced load and congestion.
		In total, data from 16\,000 jobs is shown.
		(E) \Gls{icmp} round-trip latency between cluster node and experiment setup under varying sustained loads within one drawer (\,* measured setup is loaded).
		Per condition, the distribution of 10\,000 recorded samples is shown;
		jitter is illustrated through the 99.9th percentile (orange crosses).
		(F) Bandwidth empirical cumulative distribution function (eCDF) between cluster node and experiment setup under varying loads (1--4 units).
		Per trace, 300 measurements with 60\,s runtime each are shown.
		For maximum packet size, we only show the measurement for four loaded units.
		(G) Number of stable deployments of either \gls{fpga} bitfiles or software releases per week.
		Releases are pooled for visualization only.
		(H) Line coverage across the full software stack for different levels of CI integration:
		software tests (sw) are always included; we additionally run test suites against a simulated \gls{fpga} and \gls{asic} (sim) and on the physical system (hw), cf.~\cref{sec:continuous-integration}.
	}
	\label{fig:measurements}
\end{figure*}

We quantify the effectiveness of this workflow by the proxy of code coverage across three different test targets:
software-only tests (baseline), software-defined tests against a live \gls{rtl} simulation of the hardware (step~\ref{cistep:simtests}), and software-defined tests against a physical hardware instance (step~\ref{cistep:hwtests}).
For each of these cases, and their combinations, we measure the line coverage of the \gls{bss2} software stack during test execution and find a substantial increase from \SI{57}{\percent} to \SI{79}{\percent} (\cref{fig:measurements}H).
Crucially, this metric understates the effectiveness of the proposed methodology, as changes to \gls{rtl} code are verified, but not included in the coverage metric.

This tight integration of the \gls{bss2} hardware platform into the test and development workflow ensures stable operation.
Crucially, the standardization of this process also enables efficient collaboration among researchers throughout all levels of seniority and substantially contributes to the operational sustainability of the platform.

We have acquired metrics for characterizing this workflow over the course of six years, starting July 2020, and show the results in \cref{tab:dora}, with the parameter selection inspired by the established DORA metric collection~\parencite{forsgren2018accelerate}.
The data include all core hardware and software design repositories that use the presented \gls{cicd} methodology.
The reported CI time is the sum over all jobs on hardware systems and can therefore exceed \SI{24}{\hour\per\day}.
Here, lead time per feature is measured as the time difference between the upload of its very first commit and the submission into production.
We classify a deployment as \emph{failed} if it was reverted by a subsequent one to mitigate an issue.
Note that this may include non-critical failures: the reported time to recovery measures the time between submission of any faulty change and its revert.
It does not equal system downtime.

\Cref{fig:measurements} shows distributions and time series data for these metrics.
As is common, both lead and recovery time distributions are right-skewed:
most changes---and fixes---are deployed much quicker than suggested by their respective mean.
We find the median lead time to be \SI{8.0}{\day}, with \SI{58}{\percent} of changes deployed within two weeks and \SI{70}{\percent} within \SI{30}{\day}, while the median recovery time is \SI{6.5}{\day}, with \num{14} out of \num{16} rollbacks occurring within \SI{30}{\day}.

Since previous lab systems did not allow capturing data with similar maturity, we cannot provide a meaningful reference for comparison of the workflow effectiveness.
However, with respect to the reference values given by \textcite{forsgren2018accelerate}, we characterize the presented methodology as \emph{slow and reliable}:
The long lead times for both feature implementation and potential rollback require us to ensure a very low failure rate.
Tightly integrating the custom hardware systems into the \gls{cicd} flow lets us achieve this goal, while still allowing for high deployment throughput.
 \section{Operational Monitoring}\label{sec:monitoring}

The health of the \gls{bss2} platform is continuously monitored on multiple levels ranging from high-level experiment fidelity through network availability to power supply status.
We collect any time series data in a Graphite database~\parencite{graphite}, event-like data in InfluxDB~\parencite{influxdb}, log files in Loki~\parencite{loki}, and use Grafana~\parencite{grafana} for visualization and alerting.
In addition, system operators annotate events, such as power outages or maintenance, directly within the monitoring data.

We collect information about the system state on multiple timescales:
Nightly Jenkins jobs assess the fidelity of experiments, such as the training of \glspl{snn} for classification tasks.
Similarly, we run bihourly high-level health checks on all unoccupied systems.
These validate the overall system state, including tests for \gls{sram} access, high-speed communication channels, and \gls{asic} power supplies.
In addition, we read out and store the currently configured \gls{fpga} design revision for future reference.

\begin{table}[b!]
	\renewcommand{\arraystretch}{1.3}
	\caption{CI/CD statistics during a six-year interval, starting July 2020.
	}
	\label{tab:dora}
	\centering
	\begin{tabular}{@{}ll@{}}
		\toprule
		contributing individuals            & \num{43}                 \\
		mean number of CI runs on \gls{bss2}  & \SI{198}{\per\day}       \\
		mean CI time on \gls{bss2}            & \SI{23.5}{\hour\per\day} \\
		total deployments (production/stable) & \num{3718}               \\
		total rollbacks                       & \num{16}                 \\
		mean deployment frequency             & \SI{1.70}{\per\day}      \\
		change failure rate (rollbacks)       & \SI{0.43}{\percent}      \\
		median lead time                      & \SI{8.0}{\day}           \\
		median time to recovery               & \SI{6.5}{\day}           \\
		\bottomrule
	\end{tabular}
\end{table}
 
Data that has to be available on a finer timescale is collected using dedicated monitoring services:
Once every minute, we execute \gls{icmp} echo and \gls{arp} requests from a central host to the \glspl{fpga} of all \gls{bss2} systems and evaluate the results in terms of response ratio and time.
Physical monitoring data, specifically supply voltages and system temperature, is measured by the system controller introduced in \cref{sec:systems} every second and downsampled to one-minute intervals for storage.

Quantities that need to be observed on shorter timescales, such as the \gls{asic}'s power consumption, are available directly within the experiment result structures.
As they are exclusively allocated to the respective user, they are not considered to be part of operational monitoring.

In summary, high-level experiment executions ensure that the neuromorphic accelerators can be calibrated and trained to fulfill a given task.
This checks both the fidelity of the analog circuits and of changes to the software algorithms that were deployed previously.
Network connectivity checks and global log aggregation provide notifications for inaccessible systems to maintainers without user interaction and generate an overview of the global system state.
On the lowest level, supply and temperature monitoring enables action already when pre-fail conditions are met.

This multi-level approach supports fault detection and diagnosis through notifications of failures and degradation.
The collected data has proven vital for understanding unexpected system behavior by allowing for comparison with the past.
We therefore set a retention period for high-resolution data of at least one year, resulting in a total storage consumption of approximately \SI{9}{\gibi\byte\per\year}.
 \section{Lessons learned}\label{sec:lessons-learned}

The presented system represents the current iteration of a decade-long endeavor to provide a reliable platform for neuromorphic computing from within a university-based research initiative.
In the following, we share lessons we have learned during this process.
While we anchor these to concrete, system-specific experiences, we believe that each point is of general applicability---although their individual relevance to other environments may vary.

\lesson{Strict resource management}
\Gls{bss2} systems can be used concurrently by different users in an interleaved fashion, with individual runtimes ranging from seconds to days.
Their \glspl{fpga} hold no session information and terminate possibly active connections when receiving a transport layer initialization request.
All user-accessible devices must therefore be managed, and any interference with unallocated resources must be prevented through automation.
\label{feat:firewall}
We do so by integrating all available hardware resources into the SLURM resource manager~\parencite{yoo2003slurm} and dynamically adjust firewall rules upon system allocation.
Without an active allocation, no packet will be routed to the respective system.

\Cref{fig:measurements}D characterizes the startup overhead introduced by these measures by plotting the total SLURM prolog runtime, during which they run.
We correlate a total of \num{16000} jobs allocating \gls{bss2} systems with the number of concurrently starting jobs on the same host, a proxy for prolog-induced load and congestion.
We find the median prolog runtime as an upper bound for our custom additions to be \SI{0.34}{\second}.

\lesson{Independence of hardware resources}
In line with the strict resource management through software mentioned above, hardware should minimize ways in which two independent systems can influence each other (e.g., through adjustments of a shared power supply).
We thereby ensure that any experiment data stream can only impact allocated resources.
This is especially relevant for public access, which is limited to specific systems and can therefore never influence the full platform.
\label{feat:supply-independence}

\lesson{Network separation}
While Ethernet has proven to be a very robust protocol for our use case, unintended behavior of a single device may stall operation through denial-of-service.
For example, faulty \gls{fpga} configurations have sent out \gls{arp} requests at wire speed in the past and thereby stalled the whole network.
For such cases, sufficient bandwidth must be available for management traffic to shut down the misbehaving device.
While separate physical networks are ideal for this purpose, they pose significant financial and spatial constraints.
We therefore limit broadcast traffic and utilize weighted scheduling policies in our switch infrastructure to guarantee management access (IEEE~\mbox{802.1Qaz}, cf.~\cref{sec:network}).

\begin{table*}[t!]
	\newcommand{\noinfo}{---}
	\newcommand{\thor}{\lightning}
	\newcommand{\ebrains}{\textepsilon}
	\renewcommand{\arraystretch}{1.3}
	\caption{Comparison of various novel computing systems}
	\label{tab:comparison}

	\begin{center}
	\begin{tabular}{@{}lllllllllllll@{}}
		\toprule
		system                                                                             & paradigm     & domain     & type/size    & \begin{minipage}{12mm}remote\\ access\end{minipage}  & \begin{minipage}{14mm}remote\\ management\end{minipage} &\begin{minipage}{14mm}system health\\ monitoring\end{minipage} & CI/CD target                                                                    \\ \midrule
		Cortical Labs CL1~\parencite{kagan2025cl1,cuthrell2025human,corticallabs2026cloud} & biological   & commercial & rack drawer  & paywall                                              & \noinfo                                                 & yes                                                           & \noinfo                                                                         \\
		Q.ANT NPS\,2~\parencite{quant2025native}                                           & photonic     & commercial & rack drawer  & local                                                & yes                                                     & \noinfo                                                       & \noinfo                                                                         \\
		\emph{n.n.}~\parencite{pogorelov2021compact}                                       & ion trap     & academic   & multi-rack   & planned                                              & yes                                                     & yes                                                           & \noinfo                                                                         \\
		Intel Loihi 1~\parencite{davies2018loihi,davies2021advancing}                      & CMOS digital & commercial & edge         & restricted                                           & \noinfo                                                 & \noinfo                                                       & \noinfo                                                                         \\
		Intel Hala Point~\parencite{intel2024hala}                                         & CMOS digital & commercial & rack drawer  & \noinfo                                              & \noinfo                                                 & \noinfo                                                       & \noinfo                                                                         \\
		SpiNNaker-1~\parencite{Khan2008,spinnaker2026integrationtests,hbp_sga1_d11-3-3}    & CMOS digital & academic   & multi-rack   & public\textsuperscript{\ebrains}                     & yes                                                     & yes                                                           & software, hardware                                                              \\
		SpiNNaker-2~\parencite{gonzales2024spinnaker,spinnaker2026spinnman2}               & CMOS digital & mixed      & multi-rack   & restricted\textsuperscript{\thor}                    & yes                                                     & \noinfo                                                       & software, hardware                                                              \\
		Speck~\parencite{yao2024spike,daiedge2026engines,synsense2022speckdatasheet}       & CMOS digital & mixed      & edge         & planned                                              & no                                                      & \noinfo                                                       & \noinfo                                                                         \\
		Anabrid Redac~\parencite{anabrid2025redac}                                         & CMOS analog  & commercial & rack cabinet & restricted                                           & yes                                                     & yes                                                           & \noinfo                                                                         \\
		DYNAPs~\parencite{moradi2018dynaps,inilabs2017dynapse}                             & CMOS analog  & academic   & tabletop     & no                                                   & no                                                      & no                                                            & \noinfo                                                                         \\
		BrainScaleS-1~\parencite{schmidt2023commissioning}                                 & CMOS analog  & academic   & multi-rack   & public\textsuperscript{\ebrains}                     & yes                                                     & yes                                                           & software                                                                        \\
		\gls{bss2} [this work]                                                             & CMOS analog  & academic   & multi-rack   & public\textsuperscript{\ebrains,\thor}               & yes                                                     & yes                                                           & \begin{minipage}[t]{25mm}software, hardware,\\ circuit simulation\end{minipage} \\
		\bottomrule
	\end{tabular}
	\end{center}

	{\scriptsize \hspace{4.8mm} \begin{minipage}[t]{16.5cm}
		\enquote{\noinfo} indicates missing public information\hspace{8mm}
		\textsuperscript{\ebrains}\,available via EBRAINS (\url{https://ebrains.eu/nmc})\hspace{8mm}
		\textsuperscript{\thor}\,available via THOR (\url{https://neuromorphiccommons.com})

		\vspace{0.5em}
		Remote access indicates whether users can execute workloads without physical access to the system through a centralized service.
		Remote management describes if maintenance is possible without physical access.
		Health monitoring is assigned if there is documented evidence of centralized infrastructure to assess system health.
		The \gls{cicd} target describes what kinds of test targets are integrated into a system's \gls{cicd} pipeline.
	\end{minipage}}
\end{table*}

\lesson{Prefer Ethernet over USB}
\label{feat:no-usb}
Previous iterations of the \gls{bss2} platform utilized deep trees of USB devices for managing the individual systems.
In practice, eventual failures of the host-side \gls{usb} subsystem or malfunctioning endpoints required frequent manual interaction to restore system availability.
Since then, we have migrated to Ethernet where possible---specifically, for the \gls{soc}-based system controller described in \cref{sec:systems}---and also prefer this protocol for new measurement equipment:
In contrast to USB, Ethernet poses fewer constraints on tree size and topology~\parencite{usb2spec} and allows galvanic isolation between devices.
USB additionally requires a fixed pairing of devices to a single host system, which in practice poses strong requirements on the locality of nodes.
Ethernet-based networking, on the contrary, allows multiple---remote---hosts to access the same equipment and thereby enables fault tolerance and load balancing.

\lesson{Monitoring with long retention periods}
A reliable monitoring infrastructure is crucial for any sustainable platform operation.
Here, we'd like to additionally highlight that retention periods of high-resolution data must be longer than it takes users to notice typical failure cases.
For the specific application described in this manuscript, the intrinsic stochasticity of most experiments can lead to multi-month periods in which users accumulate errors until they report them.

\lesson{Physical access control requires remote management}
Even in a university-based context, physical access to production systems must be limited to a small group of maintainers.
While this statement itself is trivial, it creates additional constraints on the system, especially for research platforms:
Past setups required---and also allowed---students and researchers to physically interact with the neuromorphic system, e.g., to attach external measurement devices.
While this enabled great flexibility in the experimental setups, it came with frequent service disruption, e.g., due to unintentionally unplugged cables.
For current-generation systems, we require all functionality to be remotely controllable, including controlled access to different reset domains, power management, and data acquisition.
This allows most systems to be operated in a hall with tight physical access restrictions, while very few are used as---potentially unstable---laboratory setups.
Due to the network infrastructure depicted in \cref{fig:network}, this split between \enquote{machine hall} and \enquote{laboratory} is transparent to the operator.

\lesson{Transparency mitigates personnel turnover}
An additional constraint specific to our academic context is a high turnover rate for workers, mostly due to students finishing their studies and leaving thereafter.
At the same time, technical documentation is sparse, as scientific advancements are prioritized over commodity tasks.
Therefore, past endeavors to comprehensively document the large and complex \gls{bss} platforms were not successful.
Even if certain components, like an \gls{asic}, an \gls{fpga} design, or a software library, were described in detail, these documents would soon become outdated.
In the absence of better alternatives, we therefore rely on at least one permanent staff scientist per team who stays with the project for extended periods of time.
These need to keep track of all ongoing developments, advancements, and architectural changes.
This task has been significantly simplified by a focus on transparency in communication and development---facilitated by the introduction of the presented open development workflow and its integration into a team-wide chat system.
Together, this ensures a good overview of all changes to the system and forces deployments to be automatically executed in well-defined pipelines---a form of executable documentation.
 \section{Discussion}\label{sec:discussion}

We have shown how we transformed a laboratory system for exploring novel compute into a publicly available research platform operated in an academic environment.

To scope this work, we have defined a set of attributes associated with such laboratory systems and established, as part of their transformation, the adoption of known best practices for the operation of such a platform:
The final system is housed in standard 19-inch rack cabinets, connected solely via Ethernet, continuously monitored and tightly integrated into a mature \gls{cicd} workflow.
In total, we estimate the hardware cost of this transformation at \num{10000}\,USD (2023, both rack cabinets).
The maintenance effort, excluding research and development, amounts to less than one \gls{fte} and is distributed across multiple scientists.

While we share several traits with conventional data centers, the presented implementation poses certain shortcomings:
Most importantly, we do not target high availability of the systems and therefore omit redundancy for most power supplies and networking.
Especially with academic personnel---instead of dedicated technical staff---operating the system, maintenance periods may be elongated.
The building itself does not come with an uninterruptible power system, making the infrastructure susceptible to power outages.\footnote{
 Since its completion in 2023, the building encountered one power cut in May 2026.
 It lasted less than \SI{1}{\hour}.
}

Other fields of novel computing have shown similar approaches of integrating laboratory systems into compute cabinets~\parencite{pogorelov2021compact,kagan2025cl1}.
In \cref{tab:comparison}, we give a comparison of multiple such systems and the maturity of their platform integration.
To the best of our knowledge, we are the only project to publish the metrics reported in \cref{tab:dora};
therefore leaving a quantitative comparison open for follow-up work.

While the \gls{cmos}-based compute substrate of \gls{bss2} does not require complex auxiliary components, its acceleration factor still poses strong constraints on network bandwidth and latency.
We therefore base our core network infrastructure on \gls{100gbe}.
It allows---due to the strong separation based on \glspl{vlan}---the described hardware-enabled CI/CD workflow to be executed on the production systems.
The infrastructure is prepared to accommodate additional scaled-up systems in active development \parencite{ilmberger2026brainscales}.

Prototypes of the presented neuromorphic system---\gls{bss2}---first became publicly available in 2017 as part of the Human Brain Project.
Since then, it underwent the described transformation process and matured to the presented platform that can be accessed via both the European research infrastructure EBRAINS and the US-based THOR project.\footnote{Access to \gls{bss2} can be requested via \url{https://ebrains.eu/nmc} and \url{https://www.neuromorphiccommons.com}.
}
Together with its predecessor system, it was moved to its current location in the \gls{einc} in 2023, with the current production system consisting of 13 \gls{bss2} setups.
Between January~2023 and September~2026, more than \num{520} individual researchers conducted an average of over \num{10800} experiments per day.
 \section*{Acknowledgments}

The authors would like to thank
Christian Mauch for his contributions to hardware resource management and the system controller,
Julian Göltz for spearheading automated system health assessment when training \glspl{snn} on the platform,
and
the Chair of Highly-Parallel VLSI Systems and Neuro-Microelectronics at Technische Universität Dresden for their contributions to \gls{bss2}.
Finally, we thank all past and present members of the Electronic Visions group for their contributions to the BrainScaleS platforms.
Generative artificial intelligence was used for copy editing of the manuscript, specifically for grammar, spelling, and language refinement.
 
\printbibliography

\end{document}